\documentclass[12pt]{JHEP3} 

\JHEPspecialurl{http://jhep.sissa.it/JOURNAL/JHEP3.tar.gz}

\usepackage{epsfig,multicol,bbm}

\title{Minimal Dark Matter and Leptogenesis}

\author{Eung Jin Chun \\
Korea Institute for Advanced Study\\
Hoegiro 87, Dongdaemun-gu, Seoul 130-722, Korea\\
Email: {ejchun@kias.re.kr}
 }

\abstract{A Dirac fermion carrying an integral weak isospin and the vanishing hypercharge is considered as its neutral component can be a promising dark matter candidate (called the minimal dark matter) whose mass is of order 100 GeV. While the symmetric population annihilates away due to a rapid gauge interaction, its asymmetric abundance is supposed to be produced by the decay of a right-handed neutrino superfield in the supersymmetric  type I seesaw mechanism.  The efficiencies for generating the dark matter and lepton asymmetries are calculated by solving a set of approximate
Boltzmann equations. A spectacular feature of this scenario is the existence of a long-lived singly- or multiply-charged scalar and a shorter-lived singly-charged fermion whose tracks can be readily looked for at the LHC.
}

\begin{document}

One of the simple and attractive ways to introduce dark matter is to postulate an extra  multiplet of the Standard Model gauge group $SU(3)_c\times SU(2)_L \times U(1)_Y$, which is called the ``minimal dark matter" (MDM) \cite{cirelli05}.
Some important features of the MDM  arise from the fact it has the usual gauge interactions.
The MDM must be a completely neutral ($T_3=Y=0$) component of a $SU(2)_L$ multiplet.
Otherwise, it should have already been observed  through its large cross-section with nuclei.  If the standard thermal freeze-out determines the cosmic abundance of the MDM, its mass should be at the multi-TeV region which is hard to be probed at the LHC. Of course, a
non-thermal production or a thermal production in a non-standard cosmology can lead to a right relic number density for a lower mass MDM.  Even in this case, various astrophysical and cosmological observations put  rather strong bounds on the MDM mass \cite{chun09}.  Non-observation of cosmic anti-proton fluxes at the PAMELA experiments limits the rate of dark matter annihilation to $W^+ W^-$ \cite{donato08} which can be interpreted as the bound: $m_{DM} > 520$ GeV. A more stringent bound,  $m_{DM}>900$ GeV, may come from the galactic center radio observation  if the dark matter distribution follows the NFW profile \cite{bertone08}.

\medskip

In this paper, we consider  leptogenesis \cite{fukugita86} in the  supersymmetric type I seesaw model \cite{davidson08} as the origin of the cosmic abundance of the MDM with the mass of order 100 GeV.
The CP violating decays of a right-handed neutrino superfield produce an appropriate asymmetric relic density in the particle and anti-particle dark matter population, which invalidates the strong astrophysical bounds mentioned above as the symmetric relic density can be sufficiently suppressed by fast gauge annihilations. The idea that the baryon and dark matter asymmetries can be produced simultaneously during the process of leptogenesis has been put forward in various contexts \cite{an09,chun10,falkowski11,haba11}. In this type of scenario, the observed ratio of the dark matter and baryon energy densities $\Omega_{DM}/\Omega_B\approx 5$ \cite{wmap} can be accounted for
by an appropriate choice of the Yukawa couplings of a heavy seesaw particle to the lepton and dark matter sectors.  A variety of other ways relating the dark matter and baryon asymmetries have been considered in the past years \cite{adm}.

 In the following, we will first construct our model superpotential extending the type I seesaw mechanism. Then, we will show how the lepton and dark matter asymmetries are generated from the decays of the lightest right-handed neutrino superfield depending on the model parameters such as the $K$ factor and the branching ratios for the lepton and dark matter sectors.  For this, the efficiency factors are computed from a set of approximate Boltzmann equations.  Finally, analyzing the mass spectrum of the scalar and fermion dark matter multiplet, the neutral fermion component will be suggested as the MDM.  This scenario provides clean signals of a long-lived charged scalar and a shorter-lived charged fermion at the LHC.

\bigskip

An important feature of our scenario is that the $B-L$ symmetry in the usual lepton sector has to be extended to the dark matter sector in a way that the particle and anti-particle dark matter carry opposite $B-L$ charges. The $B-L$ symmetry is supposed to be broken by the right-handed neutrino mass terms and thereby the lepton and dark matter asymmetries are generated from CP-violating decays of a right-handed neutrino.  Thus our dark matter candidate is a vector-like multiplet $(\Sigma, \Sigma^c)$\footnote{In the original paper \cite{cirelli05}, the MDM is a Majorana particle.} which carries the weak isospin $T=1,2,\cdots$ of $SU(2)_L$ with $Y=0$.
The superpotential term added to the Supersymmetric Standard Model sector is
\begin{equation} \label{Wmdm}
 W_{\rm new}= y_{ij} N_i L_j H_u + {1\over2} h_{ijk} N_i \Sigma_j \Sigma_k  + m_{\Sigma_i} \Sigma_i\Sigma^c_i + {1\over2} M_i N_i N_i
\end{equation}
where $i,j$ and $k$ are flavor indices for the heavy right-handed neutrino $N$ and the lepton doublet $L$, and also possibly for the dark matter multiplet $\Sigma$.
Here the $B-L$ charges are assigned as follows:
\begin{equation}
\begin{array}{c|cccc|c}
 \mbox{superfields} & L & N & \Sigma & \Sigma^c & M\cr
 \hline
 B-L & -1 & 1 & -{1\over2} & {1\over2} & -2\cr
\end{array}
\end{equation}
where also shown is the charge $-2$ of the mass parameter $M$ breaking the $B-L$ symmetry
explicitly. Note that $B-L$ can be considered as a gauge symmetry which is broken spontaneously by a vacuum expectation value of a field inducing the mass $M$ through a certain Yukawa coupling.
The first and last terms of Eq.~(\ref{Wmdm}) are the standard seesaw terms which produce
the light neutrino mass matrix:
\begin{equation}
  m^\nu_{ij}= - y_{ki}y_{kj} {\langle H^0_u\rangle^2 \over M_k} \,.
\end{equation}
For the observed neutrino masses $m_\nu \lesssim 0.1$ eV,
the sizes of the Yukawa couplings can be estimated roughly as
$y^2 \lesssim 10^{-4} (M/10^{10}\, {\rm GeV})$.

In the most part of the following discussion, we will take the triplet dark matter superfields: $\Sigma=(\Sigma^+,\Sigma^0,\Sigma^-)$ and $\Sigma^c=(\Sigma^{c+},\Sigma^{c0},\Sigma^{c-})$ as a typical example. The dark matter particle will be assumed to be a fermion component having the mass $m_{DM}=m_\Sigma$. That is, the scalar components (denoted by $\tilde{\Sigma}$ and $\tilde{\Sigma}^c$) of the triplet superfields are heavier than the fermion components (denoted also by $\Sigma$ and $\Sigma^c$).  The mass spectrum of the scalar triplets, which depends also on the soft supersymmetry breaking parameters and the D-terms,
will be discussed later.

Before considering the asymmetric MDM abundance from leptogenesis,
let us remind that the symmetric population generated by the usual thermal freeze-out is given by \cite{cirelli05}
\begin{equation}
 \Omega_{SDM}h^2 \approx 0.1 \left( 2.4\,{\rm TeV} \over m_{DM}\right)^2 \,.
\end{equation}
Thus, for the MDM with $m_{DM} \ll 1$ TeV, the symmetric component has a negligible contribution
to the dark matter density and thus its annihilation becomes hard to be observed in indirect searches of dark matter.

\bigskip

CP violating decays of $N \to L H_u$ and $\Sigma\Sigma$  generate the lepton and dark matter asymmetries.  In discussing leptogenesis,  we will take one field approximation suppressing the flavor indices of $L$ and $\Sigma$, which is enough to capture main features of our scenario.\footnote{For a more general consideration and possible flavor effects, see Ref.~\cite{davidson08}.}  That is, we consider a simple form of the right-handed neutrino Yukawa terms:
\begin{equation}
 y_i N_i L H_u + {1\over 2 } h_i N_i \Sigma\Sigma \,.
\end{equation}
The CP asymmetries of the $N_1$ decay are induced only by self-energy diagrams and take the forms of
\begin{eqnarray}
 \varepsilon_L \approx {1\over 4\pi} { \sum_i \mbox{Im}[y_i y_1^*(y_i y_1^* + h_i h_1^*)] \over |y_1|^2 + {3\over4}|h_1|^2 } {M_1\over M_i}\,, \\
 \varepsilon_{DM} \approx {2\over 4\pi} { \sum_i \mbox{Im}[h_i h_1^*(y_i y_1^* + h_i h_1^*)] \over |y_1|^2 + {3\over4} |h_1|^2 } {M_1\over M_i}\,,
\end{eqnarray}
where we assumed $M_1 \ll M_{2,3}$.
The decay rates of $N_1$ to $L H_u$ and $\Sigma\Sigma$ are
$\Gamma_L = 4|y_1|^2/16\pi$ and $\Gamma_{DM} = 3|h_1|^2/16\pi$, respectively.

\begin{figure}
\begin{center}
\includegraphics[width=0.70\linewidth]{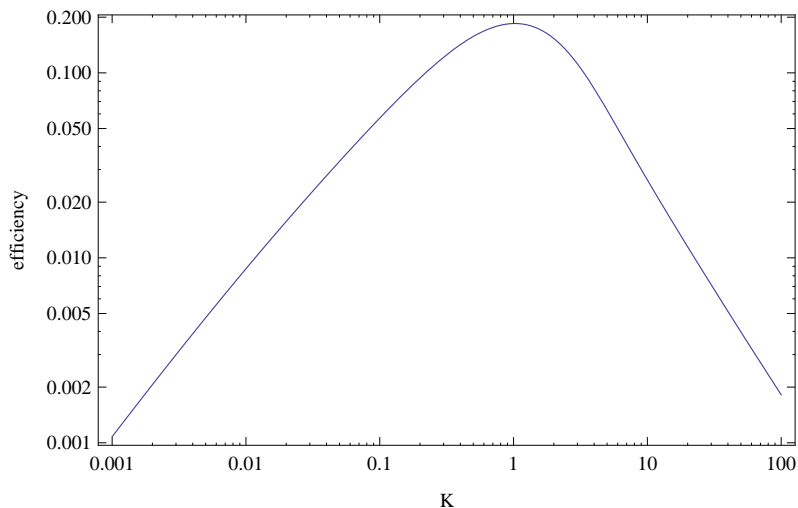}
\end{center}
\caption{The efficiency factor as a function of $K=\Gamma_{N_1}/H(T=M_1)$ derived from the approximate Boltzmann equations (10,11),
which recovers the usual leptogenesis result in the limit of $B_{DM}=0$.
}
\label{eff-K}
\end{figure}

The lepton and dark matter number asymmetries normalized by the entropy density, $Y_{L,DM}\equiv n_{L,DM}/s$, are determined by the above CP asymmetric quantities
and the efficiency factors $\eta_{L,DM}$:
\begin{equation}
 Y_{L,DM} = {315 \zeta(3)\over 4  \pi^4 g_* }\, \varepsilon_{L,DM}\, \eta_{L,DM}
\end{equation}
where $g_* \approx 250$ is the relativistic degrees of freedom including the dark matter triplets. Note that
the baryon asymmetry converted from the lepton asymmetry is  $Y_B=(10/31)Y_L$.
The efficiency factor $\eta_{L,DM}$ will depends on the branching ratio $B_{L,DM}$ for the lepton and dark matter sector, respectively, and the $K$ factor defined by
\begin{equation}
 K= {\Gamma_{N_1} \over H(T=M_{N_1})} \sim {\tilde{m}_\nu \over 10^{-3}\, \rm eV}
\end{equation}
where $\Gamma_{N_1}=\Gamma_L+\Gamma_{DM}$ and the second relation follows from
the assumption of $|y_1| > |h_1|$ with $\tilde{m}_\nu \equiv |y_1|^2 \langle H_u^0 \rangle^2/M_1$.  Note that $K\lesssim 100$ for the neutrino mass scale typically smaller than about $0.1$ eV.

In order to estimate the efficiency, we will solve the following simplified Boltzmann equations:
\begin{eqnarray}
 Y_{N_1}' &=& -z K (\gamma_D + \gamma_S) [ Y_{N_1} - Y_{N_1}^{eq}] \label{Boltz1}\\
 Y_L' &=& z K \gamma_D [\varepsilon_L  (Y_{N_1} - Y_{N_1}^{eq})
 - B_L {Y_{N_1}^{eq} \over 2 Y_l^{eq}} Y_L]   \label{Boltz2}\\
 Y_{DM}' &=& z K \gamma_D  [\varepsilon_{DM} (Y_{N_1} - Y_{N_1}^{eq})
 - B_{DM}   {Y_{N_1}^{eq} \over 2 Y_\Sigma^{eq}} Y_{DM} ] \,. \label{Boltz3}
\end{eqnarray}
 Here $\gamma_D=K_1(z)/K_2(z)$ comes from the thermally averaged decay rate, $\gamma_S$ denotes the scattering rate,  and $B_{L}$ and $B_{DM}$ are the branching ratios to the lepton and dark matter sector, respectively.  Among the scattering terms,  the $\Delta (B-L)=2$ processes are not included as they are of ${\cal O}(y^4)$ or ${\cal O}(h^4)$ and thus can be safely  neglected in the low $M$ region where  $|y|, |h|\ll1$. Our approximate calculation does not include various effects due to renormalizations, thermal corrections and gauge interactions, etc.  A more complete analysis considered in Ref.~\cite{giudice03}  can lead to ${\cal O}(1)$ changes in the final results.
The inclusion of the $\Delta(B-L)=1$ scattering effect in Eq.~(\ref{Boltz1}) is important for $K\ll1$ as it significantly enhances the $N_1$ population in the case of the vanishing initial abundance \cite{buchmuller04}.
Note  that Eqs.~(\ref{Boltz2},\ref{Boltz3}) do not have scattering terms as their effect is not essential for the degree of precision aimed in this work. We will see later that the efficiency factor in this approximation agrees reasonably with the previous result.
The $\Delta (B-L)=1$ scattering rate $\gamma_S$
comes from the $s$ or $t$ channel processes of $N_1$ and $L$ or $\Sigma$: $\gamma_S =  (2 \gamma_s^L +4 \gamma_t^L) + (2 \gamma_s^{DM} +4 \gamma_t^{DM})$.  In the following, we will work in the limit of $B_{DM} \ll B_L$ so that $\gamma_{s,t}^{DM}$ becomes sub-dominant.   Then, the analytic approximation for $\gamma_D+\gamma_S$ derived in Ref.~\cite{buchmuller04} is adopted for our calculation:
\begin{equation}
 \gamma_D + \gamma_S \approx {9\over 8 \pi^2} \left[ 1 + \ln\left(M_1\over M_h\right) z^2
 \ln\left( 1 + {a\over z}\right) \right]
\end{equation}
with $a=8\pi^2/9\ln(M_1/M_h)$ and $M_h/M_1=10^{-5}$.

\bigskip

Let us now present the solutions of the above Boltzmann equations.
First we calculate the efficiency factor $\eta_L$ as a function of $K$ in the limit of $B_{DM}=0$.  As shown in Fig.~1, our result well agrees with that of Ref.~\cite{buchmuller04}
except in the region of $K\sim1$ where a slight deviation is found.
This justifies our approximated Boltzmann equations in Eqs.~(\ref{Boltz1},\ref{Boltz2},\ref{Boltz3}).

\begin{figure}
\begin{center}
\includegraphics[width=0.49\linewidth]{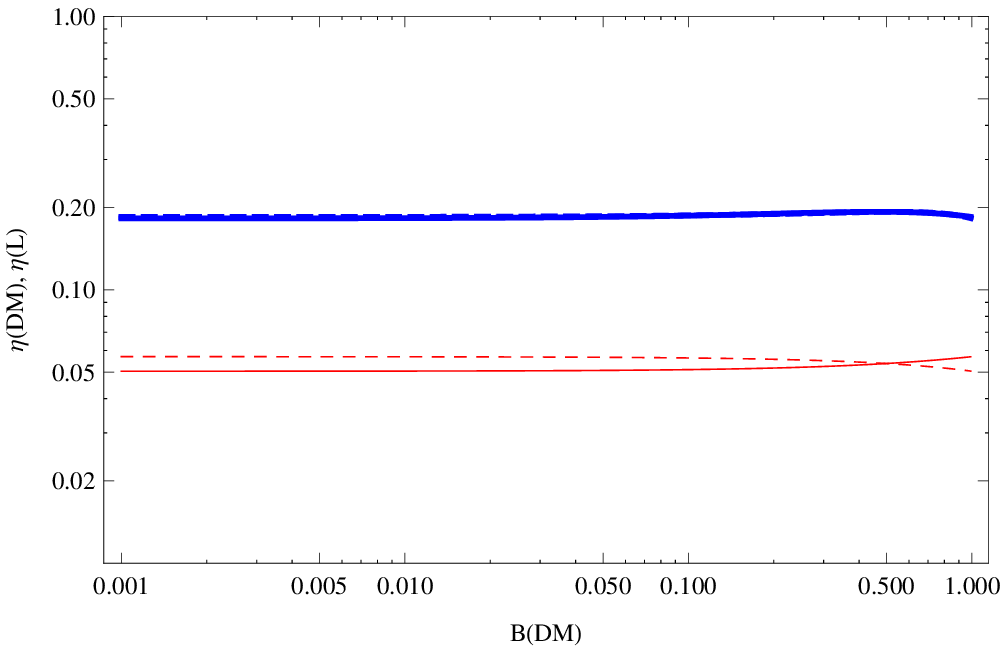}
\includegraphics[width=0.49\linewidth]{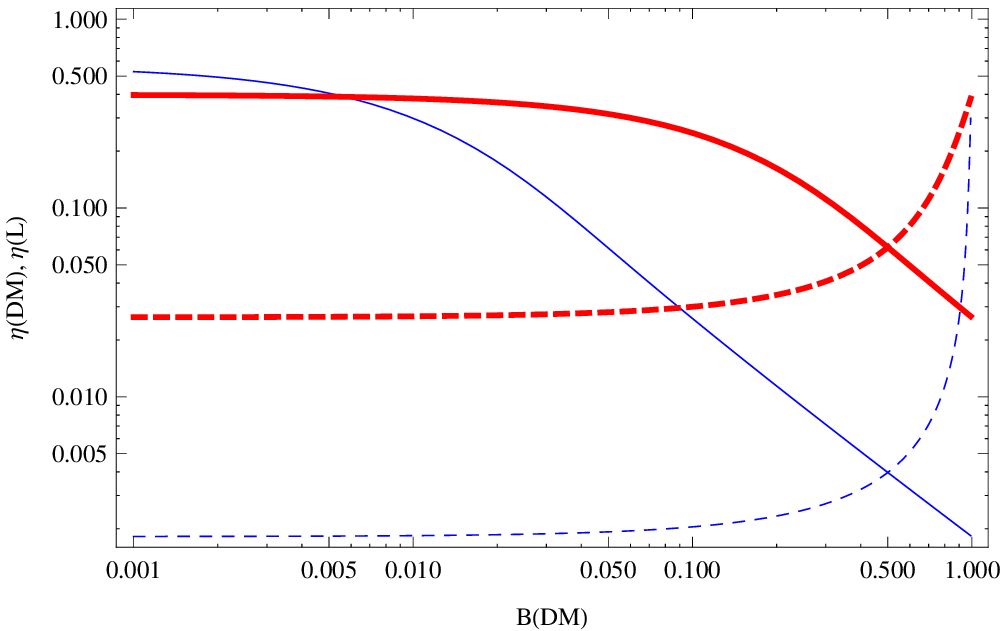}
\end{center}
\caption{The efficiency factors $\eta_{DM}$ (solid lines) and $\eta_L$ (dotted lines) for $K=0.1$ (thin red lines) and 1 (thick blue lines) in the left panel, and for $K=10$ (thick red lines) and 100 (thin blue lines) in the right panel.}
\label{eff-BDM}
\end{figure}

In Fig.~2, the efficiency factors for the lepton ($\eta_L$) and dark matter ($\eta_{DM}$) are plotted in terms of the dark matter branching ratio $B_{DM}$ for different values of $K=0.1,1,10,100$. Recall that the results for $B_{DM} \gtrsim 0.1$ are not reliable as the scattering rates involving the dark matter are not included in the Boltzmann equations. Nevertheless, the plots shows that $\eta_{DM}=\eta_{L}$ at $B_{DM}=B_{L}=1/2$ as it should be. In the left panel of Fig.~2 ($K\leq1$) one can see that the efficiencies $\eta_{DM}$ (solid lines) and $\eta_L$ (dotted lines) are almost same independently of $B_{DM}$, while both of them drop as $K$ like in Fig.~1. This  can be understood from the fact that the inverse decay terms proportional to $B_L$ and $B_{DM}$ in Eqs.~(\ref{Boltz2},\ref{Boltz3}) are both small and lead to negligible wash-out effects.  An interesting feature occurs for $K\gg1$ as shown in the right panel.  In the region of $B_{DM}\ll1$ (or $B_L\approx1$), $\eta_L$ drops as $K$ increases consistently with Fig.~1, but $\eta_{DM}$ can be even larger than the case with $K=1$. This occurs when $B_{DM} K <1$ for which the wash-out effect becomes weak.  Furthermore, $\eta_{DM}$ turns out to be larger for larger $K$ and sufficiently small $B_{DM}$. This is driven by the larger source term (the first term in Eq.~(\ref{Boltz3})).

\begin{figure}
\begin{center}
\includegraphics[width=0.70\linewidth]{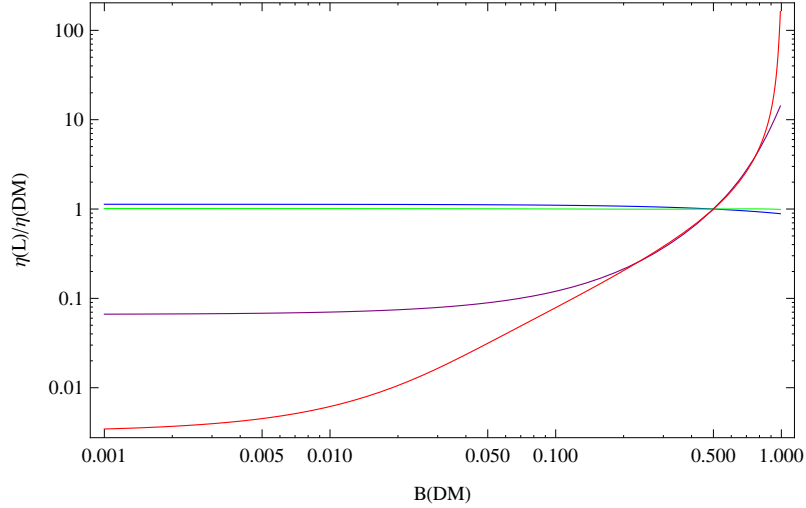}
\end{center}
\caption{
In terms of $B_{DM}$ plotted are the ratios, $\eta_{L}/\eta_{DM}$ for $K=0.1, 1, 10$ and 100 (blue, green, purple, and red solid lines from above). Note that two lines for $K=0.1$ and 1 almost overlap. The observed value of  $ \Omega_{DM}/\Omega_{B} \approx 5$ can be found for $\varepsilon_{DM}/\varepsilon_{L} \approx 8\times10^{-3}(\eta_L/\eta_{DM}) (m_{DM}/200\,\rm GeV)$.
 }
\label{sol-BDM}
\end{figure}

Fig.~3 shows
the ratio $\eta_L/\eta_{DM}$ as a function of $B_{DM}$ for $K=0.1,1,10,100$.
Given $B_{DM}$, one can now find the appropriate value of $\varepsilon_{DM}/\varepsilon_L$ satisfying the observed ratio $\Omega_{DM}/\Omega_B\approx 5$ from the relation:
\begin{equation}
 {\Omega_{DM} \over \Omega_B } = {m_{DM} Y_{DM} \over m_B Y_B} \approx {31\over10} {\varepsilon_{DM} \over \varepsilon_L} {\eta_{DM} \over\eta_L } {m_{DM} \over 1\,\rm GeV}
\end{equation}
where the prefactor $31/10$ comes from the lepton-to-baryon conversion factor.  For $m_{DM}=200$ GeV, one needs $\varepsilon_{DM}/\varepsilon_L \approx 8\times10^{-3}(\eta_L/\eta_{DM})$. This relation can be easily obtained by adjusting the Yukawa couplings $y_i$ and $h_i$. As an illustration, let us take a rough estimation of $\varepsilon_{DM}/\varepsilon_L \sim (h_i h_1)/(y_i y_1) \sim [(h_i/h_1)/(y_i/y_1)] B_{DM}$ for $B_{DM} \sim |h_1|^2/|y_1|^2 \ll 1$.  Thus, it is required to have the Yukawa hierarchies satisfying $(h_i/h_1)/(y_i/y_1) \sim $ $(8\times10^{-3}/B_{DM})$ $(\eta_L/\eta_{DM})$ $(m_{DM}/200\,\rm GeV)$.

\bigskip

So far, our discussion does not depend on whether the dark matter particle is a fermion or a scalar
component of the triplet superfield. If the MDM is a fermion, interesting collider signatures can be looked for.
The scalar components of the triplet superfields, denoted by $\tilde{\Sigma} = (\tilde{\Sigma}^+, \tilde{\Sigma}^0, \tilde{\Sigma}^-)$ and  $\tilde{\Sigma}^c=(\tilde{\Sigma}^{c+},\tilde{\Sigma}^{c0}, \tilde{\Sigma}^{c-}) $, have the mass-squared matrix in the basis of $(\tilde{\Sigma}^\lambda, (\tilde{\Sigma}^{c-\lambda})^*)$:
 \begin{equation} \label{matpm}
 {\cal M}^2 = \left[
 \begin{array}{cc}
  m_\Sigma^2+\tilde{m}^2+ \lambda  m_Z^2 c_W^2 c_{2\beta} & B m_\Sigma \cr
  B m_\Sigma & m_\Sigma^2+\tilde{m}^2-  \lambda  m_Z^2 c_W^2 c_{2\beta} \cr
 \end{array} \right]\,,
 \end{equation}
where $\lambda=\pm,0$ denotes the electric charge ($Q=T_3$), $\tilde{m}^2$ is the soft supersymmetry breaking  mass, $Bm_\Sigma$ is the soft mixing mass, and the $m_Z^2$ term comes from the  $SU(2)_L$ D-term.
For each $\lambda$, there are two mass eigenstates $\tilde{\Sigma}^\lambda_{2,1}$ whose masses are give by
\begin{equation}
 m^2_{\tilde{\Sigma}^\lambda_{2,1}} = m_{\Sigma}^2+\tilde{m}^2 \pm \sqrt{B m_\Sigma + \lambda^2 m_Z^4 c_W^4 c^2_{2\beta}} \,.
\end{equation}
Let us consider only the lighter states;  $\tilde{\Sigma}^\lambda_1$.
Note that there are two degenerate complex fields $\tilde{\Sigma}^\pm_1$ which are lighter than $\tilde{\Sigma}^0_1$.
In the limit of $Bm_\Sigma \gg m_Z^2$, their mass gap is $\Delta m \equiv m_{\tilde{\Sigma}^0_1} - m_{\tilde{\Sigma}^\pm_1}
\approx m_Z^4 c_W^4 c_{2\beta}^2/4Bm_\Sigma m_{\tilde{\Sigma}^0_1}$, which is around 1 GeV for $B= m_{\Sigma} = m_{\tilde{\Sigma}^0_1} =200$ GeV.  Recall that the electroweak radiative correction induces the mass gap
$\Delta m \approx -166$ MeV for the scalar and fermion components \cite{cirelli05}.  The former tree-level mass gap is typically larger than this radiative mass gap as far as the triplet masses and also $B$ parameter are not too high, which is the parameter region we are interested in.
Thus the dark matter component must be a neutral fermion $(\Sigma^0, \Sigma^{c0})$ resulting in the following mass hierarchy among various components of the triplet superfield:
\begin{equation} \label{mass-hierarchy}
 m_{\tilde{\Sigma}^0_1} > m_{\tilde{\Sigma}^\pm_1} > m_{\Sigma^\pm} > m_{\Sigma^0} \,.
\end{equation}
Note that $m_{\tilde{\Sigma}^0_1} - m_{\tilde{\Sigma}^\pm_1} \sim 1$ GeV and  $m_{\Sigma^\pm} - m_{\Sigma^0}
\sim 0.1$ GeV, but $m_{\tilde{\Sigma}^\pm_1}$ can be much larger than $m_{\Sigma^\pm}$.
Due to the small mass gap, $\tilde{\Sigma}^\pm_1$ and $\Sigma^\pm$ can decay  through  the off-shell
$W^\pm$ leading the decays: $\tilde{\Sigma}^0_1 \to \tilde{\Sigma}^\pm_1 \pi^\mp$ and $\Sigma^\pm \to \Sigma^0 \pi^\pm$. The corresponding decay rates are determined by the sizes of the mass gap independently of the particle masses. The decay rate of the second process is given by \cite{cirelli05}:
\begin{equation} \label{Gpi}
 \Gamma_{\pi^\pm} = T(T+1){ G_F^2 V_{ud}^2 \Delta m^3 f_\pi^2 \over \pi}
 \sqrt{1-{m_{\pi^\pm}^2\over \Delta m^2}}
\end{equation}
where we have $T=1$ for the case of the dark matter triplet. Putting the values of $\Delta m \approx 166$ MeV, $f_\pi=131$ MeV and $m_{\pi^\pm}=140$ MeV, one gets the decay length: $\Gamma^{-1}_{\pi^\pm} \approx 106$ cm. This leads to a clean signal of charged particle tracks disappearing to secondary soft pions, which can be searched for to test the idea of the MDM.

Furthermore, we can have another interesting signature coming from the scalar sector. The lightest scalar component $\tilde{\Sigma}^\pm_1$ may  decay to $\chi^0 \Sigma^\pm$ or $\chi^\pm \Sigma^0$ where $\chi^0$ and $\chi^\pm$ denote a neutralino and a chargino in the supersymmetric standard model sector, respectively. However, such decay modes are forbidden kinematically if all the three particle masses are not very different or $\chi^{0,\pm}$ are heavier than $\tilde{\Sigma}^\pm$.
In this case, $\tilde{\Sigma}^\pm$ can decay only through the exchange of the heavy right-handed neutrino. From Eq.~(\ref{Wmdm}), we have the low-energy effective superpotential
\begin{equation}
 W_{eff} = {y h\over 2 M} L H_u \Sigma \Sigma
\end{equation}
where flavor indices are suppressed for simplicity. From this, one gets the coupling $
\xi\, \nu \Sigma^\pm \tilde{\Sigma}^\mp_1$ allowing the decay:
\begin{equation}
  \tilde{\Sigma}_1^\pm \to \nu \Sigma^\pm \to \nu \pi^\pm \Sigma^0
\end{equation}
with a tiny Yukawa coupling $\xi = y h \langle H^0_2\rangle/2M$.
As a rough estimate, let us take $y\sim h$ and $m_\nu \sim y^2 \langle H_u^0 \rangle$ leading to $\xi \sim m_\nu/\langle H_u^0 \rangle \sim 10^{-12}$ for $m_\nu\sim 0.1$ eV.  Therefore, $\tilde{\Sigma}_1^\pm$ behaves like a stable charged particle which  will leave slowly-moving and highly-ionizing tracks inside detectors.

A more spectacular signature follows if the MDM is a neutral fermion component of a superfield with the weak isospin $T\geq2$.  Generalizing Eqs.~(\ref{matpm},\ref{mass-hierarchy}) for higher isospin $T$, one can see that the lightest scalar component with $|Q|=|T_3|\geq2$ can decay only to a neutrino and its charge fermion partner again through a small Yukawa coupling $\sim m_\nu/\langle H_u^0\rangle$. In the fermion sector, applying Eq.~(\ref{Gpi}) for $T=2$ and 3, we get $\Gamma^{-1}_{\pi^\pm} \approx 35$ cm and 18 cm, respectively, which are still long enough to be traced.  Note that a multiply-charged fermion decays faster due to a larger mass gap and thus its tracks are too short to be observed.
Therefore, one can look for rather short singly-charged (fermion) tracks and very long muliply-charged (boson) tracks to test the model.

\bigskip

In conclusion, we considered
a MDM with $m_{DM} \sim {\cal O}(100)$ GeV as a promising dark matter candidate
whose abundance is produced asymmetrically during the process of
leptogenesis in the supersymmetric type I seesaw mechanism.  For such a low mass ($\ll $ TeV), the symmetric population becomes much smaller than $\Omega_{DM} h^2 \approx 0.1$ and thus various strong
upper bounds from astrophysical and cosmological observations can be evaded. The amounts of the lepton/baryon and dark matter asymmetries produced by CP violating decays of a heavy right-handed neutrino depends on the CP asymmetric quantities, the efficiency factors and the branching ratios of the lepton and dark matter sector.  Basically, these quantities are controlled by the Yukawa couplings of the right-handed neutrino and the observed ratio of $\Omega_{DM}/\Omega_B$ can be easily obtained by a reasonably hierarchical Yukawa structure.

The lightest scalar component of a weak isospin multiplet superfield with $Y=0$ and $T=1,2, \cdots$
is typically a scalar field with $Q = T_3=\pm 1, \pm2, \cdots$ due to the $SU(2)_L$ D-term contribution, and thus the neutral fermion component necessarily becomes the MDM.   If the decay of such a scalar particle to the usual lightest supersymmetric particle
(such as a bino) and its fermion superpartner  is kinematically forbidden, it can decay only to a neutrino with a  tiny Yukawa coupling of order $m_\nu / \langle H_u^0 \rangle$,
and thus stable in the collider time scale. Also the singly-charged fermion companion of the MDM leaves a disappearing charged track whose length is maximally about 100 cm for $T=1$.
Thus, our asymmetric MDM scenario can be tested cleanly at the LHC experiments by the observation of these two kinds of slowly-moving and highly-ionizing tracks coming from a singly-charged fermion and a singly- or multiply-charged boson.

\bigskip

{\bf Acknowledgments:} This work was supported by Korea Neutrino
Research Center through National Research Foundation of Korea
Grant (2009-0083526).

\end{document}